\documentclass[a4paper,11pt]{article}
\usepackage{aaskaiid}
\usepackage{orcidlink}
\usepackage{physics}
\usepackage[Euler]{upgreek}
\title{Unveiling the Microhertz Gravitational-Wave Sky with the Square Kilometre Array Observatory}
\ShortTitle{Unveiling the \textmu{Hz} GW Sky with the SKAO}

\author[1,2]{Alexander C. Jenkins~\orcidlink{0000-0003-1785-5841}}
\ShortName{Alexander C. Jenkins et al.} 
\author[3,4]{Diego Blas~\orcidlink{0000-0003-2646-0112}}
\author[5]{Joshua W. Foster~\orcidlink{0000-0002-7399-2608}}

\affiliation[1]{Kavli Institute for Cosmology, University of Cambridge, Madingley Road, Cambridge CB3 0HA, UK}
\affiliation[2]{DAMTP, University of Cambridge, Wilberforce Road, Cambridge CB3 0WA, UK}
\emailAdd{acj46@cam.ac.uk}
\affiliation[3]{Institut de Fisica d'Altes Energies (IFAE), The Barcelona Institute of Science and Technology, Campus UAB, 08193 Bellaterra (Barcelona), Spain}
\affiliation[4]{Institució Catalana de Recerca i Estudis Avançats (ICREA), Passeig Lluís Companys 23, 08010 Barcelona, Spain}
\emailAdd{dblas@ifae.es}
\affiliation[5]{Department of Physics, University of Wisconsin-Madison, Madison, WI 53706, USA}
\emailAdd{jwfoster@wisc.edu}

\abstract{The gravitational-wave sky is expected to contain a rich variety of sources across a very broad range of frequencies.
Much like in electromagnetic astronomy, exploring new gravitational-wave frequency bands therefore has the potential to unlock powerful new insights into the Universe.
In this chapter, we investigate the prospects for using high-precision timing of binary millisecond pulsars with the Square Kilometre Array Observatory (SKAO) to search for gravitational waves in the microhertz (\textmu{Hz}) frequency band by targeting resonant gravitational-wave perturbations to the orbits of these binaries.
Using only a handful of known systems, we show that SKAO observations can achieve unprecedented sensitivity to microhertz gravitational waves, with the potential to detect inspiralling massive black hole binaries in this band.
These searches are complementary to conventional pulsar timing array analyses, adding a new dimension to the gravitational-wave science achievable with the SKAO.}

\begin{document}
\maketitle

\section{Introduction}
Since the first direct detection of gravitational waves (GWs) by the LIGO-Virgo Collaboration~\citep{LIGOScientific:2016aoc}, GW astronomy in the hertz to kilohertz frequency band has developed into a powerful observational tool, probing a rich variety of physical questions that are inaccessible with electromagnetic observations alone~\citep{Sathyaprakash:2009xs}.
More recently, pulsar timing arrays (PTAs) have found increasingly compelling evidence for a stochastic GW background signal at nanohertz frequencies~\citep{NANOGrav:2023gor,EPTA:2023fyk,Reardon:2023gzh,Xu:2023wog,Miles:2024seg}, opening up a second observational window in the GW frequency spectrum, and with it a new and complementary range of GW sources and physical phenomena to investigate.

Despite these successes, large swathes of the GW frequency spectrum remain completely unexplored, including many decades in frequency between the bands probed by PTAs and ground-based interferometers such as LIGO and Virgo.
This gap will be partially filled by the future space-based interferometer LISA~\citep{LISA:2024hlh}, with peak sensitivity in the millihertz band.
There are also a number of proposed experiments targeting higher frequencies than this, ranging from decihertz~\citep{Crowder:2005nr,Kawamura:2011zz,Badurina:2019hst} all the way up to gigahertz~\citep{Aggarwal:2020olq}.
However, the \textbf{microhertz (\textmu{Hz}) frequency band} has proved particularly challenging to access with established methods, requiring either extremely long-baseline space-based interferometers~\citep{Sesana:2019vho}, or extremely high-cadence and high-sensitivity PTA campaigns, both of which are infeasible for the foreseeable future.

Recent work has investigated a new method for probing GWs in this microhertz frequency gap, using the phenomenon of \textbf{binary resonance}~\citep{Hui:2012yp,Blas:2021mpc,Blas:2021mqw,Foster:2025csl,Foster:2025nzf}.
When a binary system is exposed to an incoming GW, it experiences a perturbing acceleration that causes deviations in its orbit.
If the frequency of the GW is close to an integer multiple of the binary's orbital frequency, the perturbations are resonantly enhanced and accumulate secularly over time, leading to a potentially observable signal.
As a result, high-precision observation of binaries with periods on the order of $\sim$10--100~days can be used to probe microhertz GWs with unprecedented sensitivity.
\textbf{Binary millisecond pulsars (MSPs)} are promising candidates for this method, as their orbital parameters can be tracked with high precision via pulsar timing, and there are several known systems with suitable orbital periods.
Microhertz GW searches are also being pursued using Lunar laser ranging data~\citep{Blas:2021mqw,Foster:2025csl,Foster:2025nzf} as well as proposed satellite laser ranging missions~\citep{Blas:2026gwp}, but binary MSPs offer the advantage of having orbits that are (in many cases) well-described by pure general relativity~\citep{Freire:2024adf}, with negligible tidal dissipation or other systematic effects that could obscure a GW signal.

In this short contribution, we investigate the prospects for using pulsar timing observations with the Square Kilometre Array Observatory (SKAO) to search for microhertz GWs via binary resonance.
We begin with a brief overview of the binary resonance method in Section~\ref{sec:resonance}, before presenting forecast sensitivities for a handful of particularly promising binary MSPs in Section~\ref{sec:forecasts}.
We conclude with a discussion of our results and an outlook for future work in Section~\ref{sec:discussion}.

\section{Binary resonance: a brief overview}\label{sec:resonance}
Much like the test masses in an interferometer, any two masses in a binary system experience a relative acceleration in the presence of a GW.
In the ``small antenna'' limit, where the wavelength of the GW is much larger than the size of the binary orbit, this acceleration is given by\footnote{
    Here we work in the proper detector frame, which is a local inertial frame centred on the binary's barycentre.}
    \begin{equation}
    \label{eq:gw-acceleration}
        \ddot{r}^i=\frac{1}{2}\ddot{h}^\mathrm{TT}_{ij}r^j,
    \end{equation}
    where $r^i$ is the separation vector between the two bodies, $h^\mathrm{TT}_{ij}$ is the transverse-traceless part of the linearised metric perturbation associated with the GW, and dots denote time derivatives.
For typical GW strains this acceleration is extremely small, making the \emph{instantaneous} perturbation to the binary very challenging to detect.
However, because the separation vector $r^i$ oscillates at multiples of the binary's orbital frequency $1/P_\mathrm{b}$, one can see from Eq.~\eqref{eq:gw-acceleration} that GWs at these frequencies can resonate with these oscillations and induce secular changes to the orbit that accumulate over time, drastically enhancing the prospects of detection.

While this method allows one to search for any GW signal whose frequency content overlaps with the binary's orbital harmonics, it is particularly well-suited to searching for signals that are \emph{broadband} (to maximise the probability of overlapping with many harmonics) and \emph{long-lived} (to maximise the secular accumulation of the signal).
An ideal search target is therefore the \textbf{Stochastic GW Background (SGWB)} generated by the incoherent superposition of many independent GW sources throughout cosmic time.
Such a signal is conventionally described in terms of its dimensionless energy density per logarithmic frequency interval,\footnote{
    We fix the Hubble constant to the Planck 2018 value $H_0=67.66~\mathrm{km\,s^{-1}\,Mpc^{-1}}$~\citep{Planck:2018vyg}.}
    \begin{equation}
    \label{eq:omega-gw}
        \Omega_\mathrm{gw}(f)=\frac{1}{\rho_\mathrm{c}}\dv{\rho_\mathrm{gw}}{\ln f}=\frac{2\uppi^2}{3H_0^2}f^2h_\mathrm{c}^2(f),
    \end{equation}
    where $\rho_\mathrm{gw}$ is the energy density in GWs, $\rho_\mathrm{c}=3H_0^2/(8\uppi G)$ is the critical energy density of the Universe, and $h_\mathrm{c}$ is the dimensionless characteristic strain (which is often used in the PTA literature in place of $\Omega_\mathrm{gw}$).

PTAs have already found strong evidence for a SGWB at nanohertz frequencies~\citep{NANOGrav:2023gor,EPTA:2023fyk,Reardon:2023gzh,Xu:2023wog,Miles:2024seg}, for which the leading candidate source is a population of inspiralling massive black hole binaries (MBHBs).
This signal is expected to extend through the microhertz frequency band and into the LISA band, with higher frequencies corresponding to MBHBs that are closer to merger.
This provides a compelling target for binary resonance searches for microhertz GWs, complementing PTA and LISA observations by providing an independent probe of the cosmological population of MBHBs in this otherwise-inaccessible frequency band.

In the presence of a SGWB signal, a binary MSP will experience a stochastic perturbing acceleration, causing its orbital parameters to undergo a random walk and diffuse through parameter space.
Observationally, this manifests as an excess contribution to the covariance of the pulse Times of Arrival (ToAs), due to these stochastic deviations between the orbit and the timing model.\footnote{
    There is also a mean offset to the orbital parameters induced by the SGWB, but this is subdominant on observational timescales~\citep{Blas:2021mpc}.}
Recent work has developed a theoretical framework to efficiently compute this excess covariance in a fully time-resolved manner~\citep{Foster:2025csl,Foster:2025nzf} at the level of the individual ToAs, giving orders of magnitude better sensitivity compared to previous analyses that relied on secular averages of the perturbations~\citep{Blas:2021mpc,Blas:2021mqw}.
In Section~\ref{sec:forecasts}, we use this new framework to forecast the sensitivity of several known binary MSPs to a SGWB signal, assuming timing precision achievable with the SKAO.

\begin{figure}[t!]
    \centering
    \includegraphics{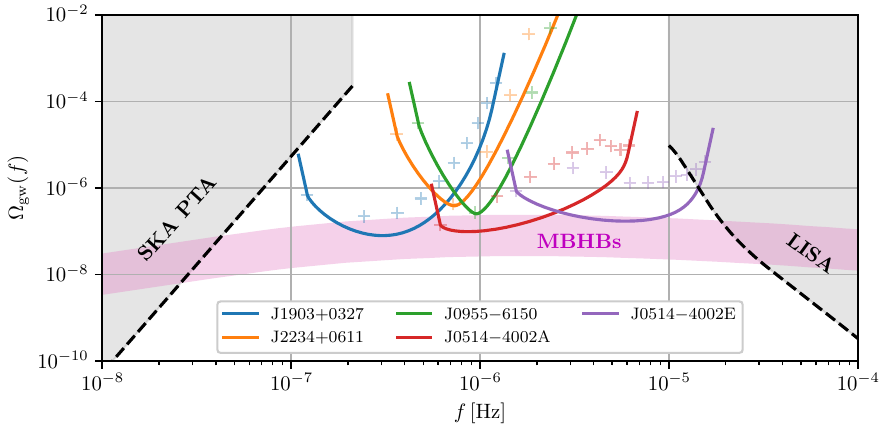}
    \caption{\label{fig:forecasts}%
    Forecast sensitivity to a stochastic gravitational-wave background in the microhertz frequency band, using resonant searches with several known binary pulsars.
    We assume a 15-year timing campaign with the SKAO at a cadence of two observations per week, with rms timing noise of 100~ns (1~\textmu{s} for J0514$-$4002A and J0514$-$4002E).
    For each binary, we show monochromatic sensitivities at the first ten harmonic frequencies (coloured points), as well as a combined power-law integrated sensitivity (solid coloured curves).
    For comparison, we also show the anticipated sensitivities of LISA~\citep{LISA:2024hlh} and of PTA searches with the SKAO~\citep{Janssen:2014dka} (black dashed curves and shaded regions).
    The pink shaded band indicates a possible signal from inspiralling massive black hole binaries~\citep{Ellis:2023owy}.}
\end{figure}

\section{Forecast sensitivity with the SKAO}\label{sec:forecasts}
Unlike PTA searches, which derive their statistical power by cross-correlating timing data from many pulsars, each system used in a binary resonance search acts as an independent detector targeting a different set of GW frequencies.
As a result, it is possible to cover the entire microhertz frequency band using only a handful of suitable binary MSPs.
There are three key criteria for selecting suitable binaries for this purpose:
\begin{itemize}
    \item \textbf{Orbital period}: The binary's orbital period $P_\mathrm{b}$ determines the set of harmonic frequencies at which it is sensitive to GWs, $(f=n/P_\mathrm{b}$ with $n=1,2,\ldots$).
    To probe the microhertz band, we therefore require binaries with periods on the order of $\sim$10--100~days.
    \item \textbf{Eccentricity}: The strength of the resonant response at each harmonic depends strongly on the binary's eccentricity $e$.
    Near-circular binaries ($e\approx0$) only respond at the second harmonic ($n=2$), while eccentric binaries ($e\gtrsim0.1$) respond at many harmonics, allowing them to probe a broader range of GW frequencies.
    \item \textbf{Timing precision}: The sensitivity of the search depends on how precisely the binary's orbital parameters can be measured via pulsar timing.
    We therefore require binaries that are well-timed, and whose orbits can be deterministically modelled to high precision.
\end{itemize}

Using these criteria, we identify five known binary MSPs that are particularly promising candidates for microhertz GW searches with the SKAO, which are listed in Table~\ref{tab:parameters}.
Of these five, two (J1903$+$0327 and J2234$+$0611) are included in the NANOGrav PTA~\citep{NANOGrav:2023hde}, and another (J0955$-$6150) is included in the MeerKAT PTA~\citep{Miles:2024rjc}.
The five systems all have significant orbital eccentricities ($e\gtrsim0.1$), and cover a range of orbital periods from $\sim$7.4 to $\sim$95~days (orbital frequencies from $\sim$1.6 to $\sim$0.12~\textmu{Hz}), allowing them to collectively span the full breadth of the microhertz band once higher harmonics are included.

\begin{table}[b!]
    \centering
    \makebox[\textwidth][c]{
    \begin{tabular}{l l l l l l l l}
        \hline\hline
        Binary pulsar & $P_\mathrm{b}\,[\textrm{day}]$ & $e$ & $\iota\,[\textrm{deg}]$ & $\omega\,[\textrm{deg}]$ & $m_\mathrm{p}\,[M_\odot]$ & $m_\mathrm{c}\,[M_\odot]$ & Reference \\
        \hline
        J1903$+$0327 & 95.174 & 0.43668 & \phantom{1}77.47 & 141.65 & 1.667 & 1.029 & \cite{Freire:2010tf} \\
        J2234$+$0611 & 32.001 & 0.12927 & 138.11 & 277.17 & 1.38 & 0.300 & \cite{Stovall:2018rvy} \\
        J0955$-$6150 & 24.578 & 0.11751 & \phantom{1}83.2 & 202.93 & 1.71 & 0.254 & \cite{Serylak:2022kna} \\
        J0514$-$4002A & 18.785 & 0.88798 & \phantom{1}63.0 & \phantom{1}82.548 & 1.39 & 1.08 & \cite{Dutta:2025flw} \\
        J0514$-$4002E & \phantom{1}7.4479 & 0.70793 & \phantom{1}52.0 & \phantom{1}65.317 & 1.53 & 2.35 & \cite{Barr:2024wwl} \\
        \hline\hline
    \end{tabular}
    }
    \caption{\label{tab:parameters}%
    Parameters of the five binary MSPs used in our forecasts: the orbital period $P_\mathrm{b}$, eccentricity $e$, inclination $\iota$, argument of pericentre $\omega$, pulsar mass $m_\mathrm{p}$, and companion mass $m_\mathrm{c}$.}
\end{table}

We investigate the sensitivity of each of these five systems to a SGWB signal using a Fisher forecast, in which we model the covariance of the pulse ToA residuals $\updelta t_a$ ($a=1,2,\ldots$) as the sum of a white-noise contribution and a SGWB-induced contribution,
    \begin{equation}
    \label{eq:covariance}
        \updelta t_a\sim\mathcal{N}(\vb*0,\vb*\Sigma),\qquad[\vb*\Sigma]_{ab}=\sigma_\mathrm{w}^2\delta_{ab}+\Omega_\mathrm{ref}[\vb*\Sigma^{(\alpha)}_\mathrm{gw}]_{ab}.
    \end{equation}
Here, $\sigma_\mathrm{w}$ is the white-noise timing precision, $\Omega_\mathrm{ref}$ is the amplitude of the SGWB at a reference frequency, which we choose to be the first harmonic for each system ($f_\mathrm{ref}=1/P_\mathrm{b}$), and $\vb*\Sigma^{(\alpha)}_\mathrm{gw}$ is the SGWB-induced covariance matrix for a power-law SGWB with spectral index $\alpha$ (defined such that $\Omega_\mathrm{gw}(f)=\Omega_\mathrm{ref}\times(f/f_\mathrm{ref})^\alpha$).
We calculate the latter using the formalism developed in \citet{Foster:2025csl,Foster:2025nzf}, which accounts for the full time dependence of the GW-induced perturbations in all six of the binary's orbital parameters.
These perturbations are then projected onto the ToA residuals using a timing model---here we consider only the leading-order Roemer delay contribution, but more sophisticated timing models (e.g. including Shapiro and Einstein delays) can be straightforwardly incorporated, and may give enhanced sensitivity.

Equation~\eqref{eq:covariance} allows us to compute the expected sensitivity to the SGWB amplitude $\Omega_\mathrm{ref}$ using the Fisher information,
    \begin{equation}
        \sigma_{\Omega}^{-2}=\frac{1}{2\sigma_\mathrm{w}^4}\Tr[(\vb*\Sigma^{(\alpha)}_\mathrm{gw})^2],
    \end{equation}
    with the corresponding 95\% confidence upper limit given by $\Omega_\mathrm{ref}^\mathrm{UL}\approx1.96\,\sigma_{\Omega}$.
By repeating this calculation for a range of spectral indices $\alpha$, we construct a power-law integrated sensitivity curve~\citep{Thrane:2013oya} for each binary system, which indicates the detection threshold for SGWB signals with sufficiently smoothly-varying spectra.

The timing white noise $\sigma_\mathrm{w}$ depends on a variety of factors, including the pulsar's intrinsic brightness and pulse profile stability, the telescope sensitivity, and the integration time used to produce each ToA.
In practice, most timing measurements with SKA-Mid will likely be dominated by pulse jitter noise~\citep{Gitika:2025nqe}, which places a floor on the achievable timing noise regardless of telescope sensitivity.
We assume a conservative jitter-limited timing precision of $\sigma_\mathrm{w}=100\,\mathrm{ns}$, while noting that significantly lower jitter levels have been measured for some MSPs~\citep{Janssen:2014dka,Gitika:2025nqe}.
For the MSPs J0514$-$4002A and J0514$-$4002E we assume an order-of-magnitude larger timing noise of $\sigma_\mathrm{w}=1\,\textrm{\textmu s}$, as both systems are belived to possess a third-body companion~\citep{Dutta:2025flw,Barr:2024wwl} and may therefore suffer from systematic timing uncertainties.
We assume a cadence of two timing observations per week for each system over a total timespan of $T_\mathrm{obs}=15~\mathrm{yr}$.
The total observing time required for such a campaign can be minimised by exploiting SKA-Mid's sub-array capability, which should allow jitter-limited timing of multiple MSPs simultaneously.
Overall, these assumptions are consistent with those used in previous SKAO PTA forecasts~\citep{Janssen:2014dka}, with the higher observing cadence balanced out by the much smaller number of MSPs included in the binary resonance search (five here, as opposed to $\gtrsim50$ in PTAs).

Our forecast sensitivities for each of the five binary MSPs are shown in Figure~\ref{fig:forecasts}.
For each system, we include the first ten harmonic frequencies (with individual monochromatic sensitivities shown as coloured points), which are combined to produce a power-law integrated sensitivity.
The resulting sensitivity curves cover the entirety of the microhertz band, with peak sensitivities on the order of $\Omega_\mathrm{gw}\sim10^{-7}$.
To demonstrate the discovery potential of these searches, we also show a predicted SGWB signal from inspiralling MBHBs~\citep{Ellis:2023owy}, which is within reach of our projected sensitivities.
This is just one of a plethora of potential SGWB signals in this band, including primordial cosmological backgrounds from first-order phase transitions or cosmic string networks~\citep{Caprini:2018mtu,Blas:2021mqw}.

Our results are conservative in the sense that we do not account for pre-existing timing data for these binaries; the secularly-growing nature of the binary resonance effect means that SGWB sensitivities grow steeply with the timespan of the observations ($\Omega_\mathrm{gw}\propto T_\mathrm{obs}^{-4}$), so consistently incorporating existing ToAs for these systems could unlock significant sensitivity gains.
It is also worth noting that the SKAO is expected to discover thousands of new MSPs, including hundreds of `high-quality' pulsars suitable for PTA campaigns~\citep{Keane:2014vja,Keane01.2026.SKA}.
It is therefore not unreasonable to anticipate the discovery of binary MSPs with even more favourable properties for microhertz GW searches than the known systems considered here.

\section{Discussion and outlook}\label{sec:discussion}
In this chapter, we have explored the prospects for using high-precision timing of binary MSPs with the SKAO to search for GWs in the microhertz frequency band.
These searches are complementary to PTA searches for nanohertz GWs that are also planned with the SKAO~\citep{Janssen:2014dka,Shannon01.2026.SKA}, targeting GW perturbations to the orbits of the binary pulsars themselves, rather than GW effects on the propagation of pulses across the galaxy.
Using only a handful of known binary MSPs, we show that a targeted but realistic timing campaign with the SKAO can cover the entire microhertz band between LISA and PTAs, with enough sensitivity to potentially detect the GW background signal from inspiralling MBHBs.
Most of the MSPs we consider are already tracked by existing PTAs, meaning that our proposed timing campaign could also contribute to SKAO PTA efforts, maximising the scientific return from these observations.
It is also highly possible that the SKAO will discover new binary MSPs that are even better suited to microhertz GW searches, further enhancing the prospects for detection.

An important avenue for future work is to carry out realistic binary resonance searches on simulated data, to investigate the impact of timing model degeneracies or other potential systematic effects on the sensitivity.
Some preliminary studies of these issues have been carried out in the case of Lunar Laser Ranging (LLR) searches for microhertz GWs using the Earth-Moon orbit~\citep{Foster:2025csl}, which are another promising application of the methods discussed here.
In that case, it was found that the GW signal decorrelates extremely well from almost all of the ranging model parameters---the sole exception being the Earth's tidal deformability, which can mimic a resonant GW effect.
Tidal dissipation is typically negligible for binary MSPs, particularly those with white dwarf companions~\citep{Freire:2012mg,Antoniadis:2013pzd} (which is true for three of the five systems we consider~\citep{Stovall:2018rvy,Serylak:2022kna,Dutta:2025flw}).
This gives us confidence that degeneracies will not significantly degrade the forecasts presented here, but detailed studies on a case-by-case basis are nonetheless warranted.
We also note that VLBI observations of the MSPs included in these searches could independently constrain several key parameters in the timing model~\citep{Janssen:2014dka}, further alleviating any degeneracies that may be present.

The history of electromagnetic astronomy demonstrates the inherent discovery potential of exploring new frequency bands, and there is every reason to believe that the same will be true in GW astronomy.
By analysing timing data from binary MSPs in a new way, the SKAO has the potential to probe the previously-unexplored microhertz frequency band, adding new colours to the GW sky.

\section*{Acknowledgments}\label{sec:acknowledgements}
A.C.J. was supported by the UK Engineering and Physical Sciences Research Council through a Stephen Hawking Fellowship (Grant No.~EP/U536684/1), and by a Gavin Boyle Fellowship at the Kavli Institute for Cosmology, Cambridge.
This publication is part of the grant PID2023-146686NB-C31 funded by MICIU/AEI/10.13039/501100011033/ and by FEDER, UE.
IFAE is partially funded by the CERCA program of the Generalitat de Catalunya.
Project supported by a 2024 Leonardo Grant for Scientific Research and Cultural Creation from the BBVA Foundation.
The BBVA Foundation accepts no responsibility for the opinions, statements and contents included in the project and/or the results thereof, which are entirely the responsibility of the authors.
Funded by the European Union. Views and opinions expressed are however those of the author(s) only and do not necessarily reflect those of the European Union or the European Research Council Executive Agency. Neither the European Union nor the granting authority can be held responsible for them. This work is supported by ERC grant (GravNet, ERC-2024-SyG 101167211, DOI:10.3030/101167211).
D.B. acknowledges the support from the European Research Area (ERA) via the UNDARK project of the Widening participation and spreading excellence programme (project number 101159929).

\bibliographystyle{abbrvnat-maxbibnames4}
\bibliography{chapter}

\begin{thebibliography}{38}
\providecommand{\natexlab}[1]{#1}
\providecommand{\url}[1]{\texttt{#1}}
\expandafter\ifx\csname urlstyle\endcsname\relax
  \providecommand{\doi}[1]{doi: #1}\else
  \providecommand{\doi}{doi: \begingroup \urlstyle{rm}\Url}\fi

\bibitem[Abbott et~al.(2016)]{LIGOScientific:2016aoc}
B.~P. Abbott et~al.
\newblock \emph{Phys. Rev. Lett.}, 116\penalty0 (6):\penalty0 061102, 2016.
\newblock \doi{10.1103/PhysRevLett.116.061102}.

\bibitem[Agazie et~al.(2023{\natexlab{a}})]{NANOGrav:2023gor}
G.~Agazie et~al.
\newblock \emph{Astrophys. J. Lett.}, 951\penalty0 (1):\penalty0 L8,
  2023{\natexlab{a}}.
\newblock \doi{10.3847/2041-8213/acdac6}.

\bibitem[Agazie et~al.(2023{\natexlab{b}})]{NANOGrav:2023hde}
G.~Agazie et~al.
\newblock \emph{Astrophys. J. Lett.}, 951\penalty0 (1):\penalty0 L9,
  2023{\natexlab{b}}.
\newblock \doi{10.3847/2041-8213/acda9a}.

\bibitem[Aggarwal et~al.(2021)]{Aggarwal:2020olq}
N.~Aggarwal et~al.
\newblock \emph{Living Rev. Rel.}, 24\penalty0 (1):\penalty0 4, 2021.
\newblock \doi{10.1007/s41114-021-00032-5}.

\bibitem[Aghanim et~al.(2020)]{Planck:2018vyg}
N.~Aghanim et~al.
\newblock \emph{Astron. Astrophys.}, 641:\penalty0 A6, 2020.
\newblock \doi{10.1051/0004-6361/201833910}.
\newblock [Erratum: Astron.Astrophys. 652, C4 (2021)].

\bibitem[Antoniadis et~al.(2013)]{Antoniadis:2013pzd}
J.~Antoniadis et~al.
\newblock \emph{Science}, 340:\penalty0 6131, 2013.
\newblock \doi{10.1126/science.1233232}.

\bibitem[Antoniadis et~al.(2023)]{EPTA:2023fyk}
J.~Antoniadis et~al.
\newblock \emph{Astron. Astrophys.}, 678:\penalty0 A50, 2023.
\newblock \doi{10.1051/0004-6361/202346844}.

\bibitem[Badurina et~al.(2020)]{Badurina:2019hst}
L.~Badurina et~al.
\newblock \emph{JCAP}, 05:\penalty0 011, 2020.
\newblock \doi{10.1088/1475-7516/2020/05/011}.

\bibitem[Barr et~al.(2024)]{Barr:2024wwl}
E.~D. Barr et~al.
\newblock \emph{Science}, 383\penalty0 (6680):\penalty0 275--279, 2024.
\newblock \doi{10.1126/science.adg3005}.

\bibitem[Blas and Jenkins(2022{\natexlab{a}})]{Blas:2021mpc}
D.~Blas and A.~C. Jenkins.
\newblock \emph{Phys. Rev. D}, 105\penalty0 (6):\penalty0 064021,
  2022{\natexlab{a}}.
\newblock \doi{10.1103/PhysRevD.105.064021}.

\bibitem[Blas and Jenkins(2022{\natexlab{b}})]{Blas:2021mqw}
D.~Blas and A.~C. Jenkins.
\newblock \emph{Phys. Rev. Lett.}, 128\penalty0 (10):\penalty0 101103,
  2022{\natexlab{b}}.
\newblock \doi{10.1103/PhysRevLett.128.101103}.

\bibitem[Blas et~al.(2026)]{Blas:2026gwp}
D.~Blas et~al.
\newblock 2026.
\newblock \emph{GUEST: Gravitational Universe Exploration with Satellite
  Tracking} (to appear).

\bibitem[Caprini and Figueroa(2018)]{Caprini:2018mtu}
C.~Caprini and D.~G. Figueroa.
\newblock \emph{Class. Quant. Grav.}, 35\penalty0 (16):\penalty0 163001, 2018.
\newblock \doi{10.1088/1361-6382/aac608}.

\bibitem[Colpi et~al.(2024)]{LISA:2024hlh}
M.~Colpi et~al.
\newblock February 2024.
\newblock \doi{https://doi.org/10.48550/arXiv.2402.07571}.

\bibitem[Crowder and Cornish(2005)]{Crowder:2005nr}
J.~Crowder and N.~J. Cornish.
\newblock \emph{Phys. Rev. D}, 72:\penalty0 083005, 2005.
\newblock \doi{10.1103/PhysRevD.72.083005}.

\bibitem[Dutta et~al.(2025)]{Dutta:2025flw}
A.~Dutta et~al.
\newblock \emph{Astron. Astrophys.}, 697:\penalty0 A166, 2025.
\newblock \doi{10.1051/0004-6361/202452433}.

\bibitem[Ellis et~al.(2023)Ellis, Fairbairn, H{\"u}tsi, Raidal, Urrutia,
  Vaskonen, and Veerm{\"a}e]{Ellis:2023owy}
J.~Ellis et al.
\newblock \emph{Astron. Astrophys.}, 676:\penalty0 A38, 2023.
\newblock \doi{10.1051/0004-6361/202346268}.

\bibitem[Foster et~al.(2025{\natexlab{a}})Foster, Blas, Bourgoin, Hees,
  Herrero-Valea, Jenkins, and Xue]{Foster:2025csl}
J.~W. Foster et al.
\newblock April 2025{\natexlab{a}}.
\newblock \doi{https://doi.org/10.48550/arXiv.2504.16988}.

\bibitem[Foster et~al.(2025{\natexlab{b}})Foster, Blas, Bourgoin, Hees,
  Herrero-Valea, Jenkins, and Xue]{Foster:2025nzf}
J.~W. Foster et al.
\newblock April 2025{\natexlab{b}}.
\newblock \doi{https://doi.org/10.48550/arXiv.2504.15334}.

\bibitem[Freire and Wex(2024)]{Freire:2024adf}
P.~C.~C. Freire and N.~Wex.
\newblock \emph{Living Rev. Rel.}, 27\penalty0 (1):\penalty0 5, 2024.
\newblock \doi{10.1007/s41114-024-00051-y}.

\bibitem[Freire et~al.(2012)Freire, Wex, Esposito-Farese, Verbiest, Bailes,
  Jacoby, Kramer, Stairs, Antoniadis, and Janssen]{Freire:2012mg}
P.~C.~C. Freire et al.
\newblock \emph{Mon. Not. Roy. Astron. Soc.}, 423:\penalty0 3328, 2012.
\newblock \doi{10.1111/j.1365-2966.2012.21253.x}.

\bibitem[Freire et~al.(2011)]{Freire:2010tf}
P.~C.~C. Freire et~al.
\newblock \emph{Mon. Not. Roy. Astron. Soc.}, 412:\penalty0 2763, 2011.
\newblock \doi{10.1111/j.1365-2966.2010.18109.x}.

\bibitem[Gitika et~al.(2025)Gitika, Shannon, Bailes, Reardon, Miles, Champion,
  and Grunthal]{Gitika:2025nqe}
P.~Gitika et al.
\newblock October 2025.
\newblock \doi{https://doi.org/10.48550/arXiv.2510.03139}.

\bibitem[Hui et~al.(2013)Hui, McWilliams, and Yang]{Hui:2012yp}
L.~Hui, S.~T. McWilliams, and I.-S. Yang.
\newblock \emph{Phys. Rev. D}, 87\penalty0 (8):\penalty0 084009, 2013.
\newblock \doi{10.1103/PhysRevD.87.084009}.

\bibitem[Janssen et~al.(2015)]{Janssen:2014dka}
G.~Janssen et~al.
\newblock \emph{PoS}, AASKA14:\penalty0 037, 2015.
\newblock \doi{10.22323/1.215.0037}.

\bibitem[Kawamura et~al.(2011)]{Kawamura:2011zz}
S.~Kawamura et~al.
\newblock \emph{Class. Quant. Grav.}, 28:\penalty0 094011, 2011.
\newblock \doi{10.1088/0264-9381/28/9/094011}.

\bibitem[Keane et~al.(2026)Keane, author2, author3, author4, and
  author5]{Keane01.2026.SKA}
E.~F. Keane et al.
\newblock In \emph{Advancing Astrophysics with the SKA -- II (AASKAII)}. 2026.
\newblock arXiv search: Report number AASKAII/Keane01.

\bibitem[Keane et~al.(2015)]{Keane:2014vja}
E.~F. Keane et~al.
\newblock \emph{PoS}, AASKA14:\penalty0 040, 2015.
\newblock \doi{10.22323/1.215.0040}.

\bibitem[Miles et~al.(2024{\natexlab{a}})]{Miles:2024rjc}
M.~T. Miles et~al.
\newblock \emph{Mon. Not. Roy. Astron. Soc.}, 536\penalty0 (2):\penalty0
  1467--1488, 2024{\natexlab{a}}.
\newblock \doi{10.1093/mnras/stae2572}.

\bibitem[Miles et~al.(2024{\natexlab{b}})]{Miles:2024seg}
M.~T. Miles et~al.
\newblock \emph{Mon. Not. Roy. Astron. Soc.}, 536\penalty0 (2):\penalty0
  1489--1500, 2024{\natexlab{b}}.
\newblock \doi{10.1093/mnras/stae2571}.

\bibitem[Reardon et~al.(2023)]{Reardon:2023gzh}
D.~J. Reardon et~al.
\newblock \emph{Astrophys. J. Lett.}, 951\penalty0 (1):\penalty0 L6, 2023.
\newblock \doi{10.3847/2041-8213/acdd02}.

\bibitem[Sathyaprakash and Schutz(2009)]{Sathyaprakash:2009xs}
B.~S. Sathyaprakash and B.~F. Schutz.
\newblock \emph{Living Rev. Rel.}, 12:\penalty0 2, 2009.
\newblock \doi{10.12942/lrr-2009-2}.

\bibitem[Serylak et~al.(2022)]{Serylak:2022kna}
M.~Serylak et~al.
\newblock \emph{Astron. Astrophys.}, 665:\penalty0 A53, 2022.
\newblock \doi{10.1051/0004-6361/202142670}.

\bibitem[Sesana et~al.(2021)]{Sesana:2019vho}
A.~Sesana et~al.
\newblock \emph{Exper. Astron.}, 51\penalty0 (3):\penalty0 1333--1383, 2021.
\newblock \doi{10.1007/s10686-021-09709-9}.

\bibitem[Shannon et~al.(2026)Shannon, author2, author3, author4, and
  author5]{Shannon01.2026.SKA}
R.~M. Shannon et al.
\newblock In \emph{Advancing Astrophysics with the SKA -- II (AASKAII)}. 2026.
\newblock arXiv search: Report number AASKAII/Shannon01.

\bibitem[Stovall et~al.(2019)]{Stovall:2018rvy}
K.~Stovall et~al.
\newblock \emph{Astrophys. J.}, 870\penalty0 (2):\penalty0 74, 2019.
\newblock \doi{10.3847/1538-4357/aaf37d}.

\bibitem[Thrane and Romano(2013)]{Thrane:2013oya}
E.~Thrane and J.~D. Romano.
\newblock \emph{Phys. Rev. D}, 88\penalty0 (12):\penalty0 124032, 2013.
\newblock \doi{10.1103/PhysRevD.88.124032}.

\bibitem[Xu et~al.(2023)]{Xu:2023wog}
H.~Xu et~al.
\newblock \emph{Res. Astron. Astrophys.}, 23\penalty0 (7):\penalty0 075024,
  2023.
\newblock \doi{10.1088/1674-4527/acdfa5}.

\end{thebibliography}

\end{document}